\documentclass{aa}
\usepackage[dvips]{graphicx}
\usepackage{psfig}
\def\I{IRAS\,20126+4104}

\def\WAT{H$_2$O}

\def\HM{H$_2$}

\def\kms{\mbox{km~s$^{-1}$}}

\def\mic{\mbox{$\mu$m}}

\def\S{\sin\psi}
\def\C{\cos\psi}
\def\G{\tan\theta}
\def\Sq{\sin^2\psi}
\def\Cq{\cos^2\psi}
\def\Gq{\cot^2\theta}
\def\az{\mbox{$\alpha_0$}}
\def\dz{\mbox{$\delta_0$}}
\def\dvdr{\mbox{$\frac{\d v}{\d R}$}}

\def\d{{\rm d}}

\begin{document}
\title{Water masers in the massive protostar \I: ejection and deceleration
}
\author{L. Moscadelli \inst{1}, R. Cesaroni \inst{2} \and M.J. Rioja \inst{3}}
\institute{
	   INAF, Osservatorio Astronomico di Cagliari, Loc. Poggio dei Pini, strada 54,
	   I-09012 Capoterra (Cagliari), Italy
\and
	   INAF, Osservatorio Astrofisico di Arcetri, Largo E. Fermi 5,
           I-50125 Firenze, Italy
\and
           Observatorio Astronomico Nacional (IGN), Apartado 1143,
	   E-28800 Alcal\'a de Henares (Madrid), Spain
 }
\offprints{L. Moscadelli, \email{mosca@ca.astro.it}}
\date{Received date; accepted date}

\titlerunning{\WAT\ masers in \I}
\authorrunning{Moscadelli et al.}

\abstract{
We report on the first multi-epoch, phase referenced VLBI observations
of the \WAT\ maser emission in a high-mass protostar associated with
a disk-jet system. The source under study, \I, has been extensively
investigated in a large variety of tracers, including \WAT\ maser VLBA data
acquired by us three years before the present observations. The new findings
fully confirm the interpretation proposed in our previous study, namely
that the maser spots are expanding from a common origin coincident with
the protostar. We also demonstrate that the observed 3-D velocities of the
maser spots can be fitted with a model assuming that the spots are
moving along the surface of a conical jet, with speed increasing for
increasing distance from the cone vertex. We also present the results of
single-dish monitoring of the \WAT\ maser spectra in \I. These reveal that
the peak velocity of some maser lines decreases linearly with time. We
speculate that such a deceleration could be due to braking of the shocks
from which the maser emission originates, due to mass loading at the shock
front or dissipation of the shock energy.
\keywords{Masers -- ISM: jets and outflows -- ISM: individual objects: \I}
}

\maketitle

\section{Introduction}

Several molecular species are known to exhibit maser emission from the sites
of newly born stars (see e.g. Elitzur \cite{elit}). This applies especially
to the regions where high-mass stars (i.e. stars in excess of
$\sim$8~$M_\odot$) form. In particular, the water line at 22.2~GHz,
originating from the $6_{16}\rightarrow 5_{23}$ transition, is the most
powerful emitter, attaining flux densities up to $10^6$~Jy in a
$\sim$1~\kms\ wide line. Since the emitting region (named ``spot'') of a
maser line is as small as 1~mas or less, \WAT\ masers are excellent tools
for high angular resolution and low sensitivity observations such as those
performed with very long baseline interferometry (VLBI). These
characteristics make it possible to use \WAT\ masers as ``test particles'' to
trace the velocity field of the gas in the densest and hence most obscured
portions of molecular clouds. This is of great help in studies of {\it
massive} star forming regions, which are hindered by two
problems: massive stars are much more distant (a few kpc) than low-mass stars
and are born deeply embedded in their parental cores. Both shortcomings are
overcome by \WAT\ maser observations: on the one hand, VLBI observations can
attain angular resolutions of $\sim$1~mas, corresponding to $\sim$1~au at
1~kpc; on the other hand, observations at 22~GHz may easily penetrate even the
highest column densities.

For these reasons, \WAT\ masers have been extensively used to pin-point
the location of young stellar objects (YSOs) and investigate the structure
and kinematics of the gas in their surroundings. Although some authors have
suggested that such masers might originate in circumstellar disks
(see e.g. Torrelles et al. \cite{tor1}, \cite{tor2}; Goddi et al. \cite{goddi}),
the common belief
is that they are associated with outflows (Felli et al. \cite{felli}).
In one case (Torrelles et al. \cite{tornat}), the maser spots have been
found to describe a perfect circle (to one part in a thousand), suggesting
that they could originate at the interface between a spherical stellar wind
from a deeply embedded YSO and the surrounding gas.

With this in mind, Moscadelli et al. (\cite{marilu}; hereafter MCR) performed
1~mas resolution observations with the very large baseline array (VLBA)
of the \WAT\ maser emission in the well-known massive YSO \I, which is the
best example to date of a high-mass protostar associated with a Keplerian
disk and bipolar outflow/jet (Cesaroni et al. \cite{cesa97}, \cite{cesa99},
\cite{cesa05}; Zhang et al. \cite{zhang}; Hofner et al. \cite{hofner};
Shepherd et al. \cite{shep}).
The scope of the observations was to establish if the masers were co-rotating
with the Keplerian disk or expanding with the outflow/jet. The former
possibility was ruled out by MCR, who demonstrated that
the \WAT\ maser spots are distributed along the axis of the outflow/jet
and may be adequately fitted with a conical model implying an expansion
velocity of $\sim$23~\kms. Such findings are of great
importance for a better understanding of the structure of the jet in
\I, which is known to be undergoing precession (Shepherd et al.
\cite{shep}). Consequently, to establish the current direction of the jet it
is necessary to observe it as close as possible to the powering source (see
Cesaroni et al. \cite{cesa05} for a thorough discussion of this topic),
which may be best achieved with a VLBI study of \WAT\ masers.

Notwithstanding the results obtained by MCR, their study was limited to the
l.o.s. component of the velocity of the spots. This is a serious limitation
in the case of \I\ because the jet axis lies very close to the plane of the
sky (Cesaroni et al. \cite{cesa99}), thus making an accurate estimate of the
expansion velocity very difficult. A small error on the estimate of the
inclination of the velocity vector may be reflected as a large uncertainty on
the true expansion speed. In order to overcome this problem and hence obtain
a more accurate description of the jet, we have performed new multi-epoch
observations with global VLBI, which resulted in proper motion measurements
of the \WAT\ maser spots. In the following we present the results of this
study and improve the model fit proposed by MCR to take into account the
spots' motion in the plane of the sky. Finally, we present the results of
single-dish monitoring of the \WAT\ maser emission from \I\ over a period of
$\sim$15~years and interpret these in the light of the VLBI findings.

\section{Observations and data reduction}
\label{sobs}

\I\ was observed using the Global array (including a subset of
EVN\footnote{The European VLBI Network is a joint facility of European,
Chinese, South African and other radio astronomy institutes funded by their
national research councils.} antennae plus the 10 antennae of the
VLBA\footnote{The VLBA is a facility of the NRAO, which is operated by
Associated Universities, Inc., under contract with the NSF.}) for 18 hours at
three epochs, on November 9 and 26, 2000, and on March 1, 2001. A group of
\ 6 EVN antennae (Effelsberg, Jodrell, Medicina, Noto, Onsala, and Shanghai)
took part in the observations at each epoch, whereas Mets\"ahovi and Robledo
observed only the first two epochs and the second epoch, respectively. The
observations were performed in phase-referencing mode alternating scans on
the maser target with scans of the phase-reference source J2007+4029, with a
cycle time of $\approx$1~min.
J2007+4029 is an intense quasar belonging to the ICRF (International Celestial
Reference Frame) list, with a relative separation from
the target maser source of 1\fdg5.
For the purpose of bandpass and phase calibration,
a scan of a few minutes on one of the strong, compact calibrators \ J2007+777 and
J2002+4725, was observed every \ 1--2 hours.

The stations recorded an aggregate of 16~MHz bandwidth in both   
circular polarizations, centered at the LSR velocity of $-$3.5~\kms\ 
(based upon a rest frequency of 22235.0798~MHz), using 2-bit sampling. The
correlation was made at the VLBA correlator in Socorro (New Mexico)
using 1024 spectral channels uniformly weighted, which gives  
a channel separation of \ 0.21~\kms.

The amplitude and phase calibration of the observed visibilities and the
mapping were accomplished with the NRAO AIPS package, following the same
procedure described by MCR. At each epoch, the maser phase-reference channel
was chosen to be the one with the strongest emission, whose velocity changes
slightly across the epochs (from $-$14.7~\kms\ to $-$15.5~\kms), and whose
structure consists of a single, unresolved spot. The absolute position of the
reference maser spot has been derived from the maps produced after applying
the corrections calculated using the phase-reference source data. These
corrections were obtained in two steps. First, after applying the calibrator
corrections, the visibilities of the phase-reference source were fringe
fitted to find the residual fringe rate produced both by differences in
atmospheric fluctuations between the calibrators and the source, and by
errors in the model used at the correlator.  Then, after correcting for the
residual fringe rate, the visibilities of the phase-reference source were
self-calibrated to remove any possible effect induced by extended spatial
structure.  For the first observing epoch, the absolute position of the
reference maser spot is:
\ RA(J2000) = 20$^{\rm h}$ 14$^{\rm m}$ 26$^{\rm s}$.0253,
\ Dec(J2000)= 41\degr\ 13\arcmin\ 32\farcs666, estimated to be accurate
within a few tenths of milliarcsecond.  As a check, we have also calculated
the absolute position with the reverse phase-referencing technique (i.e.
mapping the phase-reference source after applying the corrections evaluated
working with the maser data), and found fully consistent results.

Using the AIPS task ``IMAGR'', \ 1\arcsec$\times$1\arcsec \ (E $\times$ N)
\ tapered, channel-averaged maps centered on the reference spot
were produced for the velocity range from $-$28 to +3~\kms, which includes all
the emission apparent in the total power spectra. We note that, at each
epoch, the maser emission extends over a velocity range narrower than that of
the VLBA observations by MCR, for which spectral features in the velocity
interval from +3 to +15~\kms\ were also observed. The field of view covered by
the tapered maps is about twice the sky area over which water maser emission
is detected in
the MCR's maps. The bulk of maser emission is found within a
distance of $\approx$200~mas from the reference spot, with the exception of a
single feature detected at a distance of $\sim$530~mas. In order to map
the maser emission at full angular and velocity resolution, at each epoch and
for each velocity channel in the range from $-$32 to +10~\kms, we produced
\ 0\farcs4 $\times$ 0\farcs4 \ (E $\times$ N) \ naturally weighted maps
centered on the reference spot. In addition to that, smaller maps were also
created at the position and velocity of the more detached feature.  The CLEAN
beam was an elliptical Gaussian with a full width at half maximum (FWHM)
size, slightly varying from epoch to epoch, of 0.6--0.8~mas along the major
axis and \ 0.5--0.6~mas along the minor axis. In each observing epoch, the
1$\sigma$ RMS noise level on the channel maps is close to the theoretical
thermal value, 3~mJy~beam$^{-1}$, for channels where no signal is detected,
and increases to 13~mJy~beam$^{-1}$ for channels with the strongest
components.

Every channel map was searched for emission above a conservative detection
threshold (in the range 5--15~$\sigma$), and the detected maser spots
were fitted with two-dimensional elliptical Gaussians, determining position,
flux density, and FWHM size of the emission. Hereafter, we use the term
``spot'' to indicate the mean position of a collection of spectrally and
spatially contiguous maser spots. As the typical line width of the water
masers is greater than 0.5~\kms, a maser feature is considered real if it is
detected in at least three contiguous channels (0.21~\kms\ wide) at the same
position, within an uncertainty equal to the FWHM obtained with the Gaussian
fit.

The uncertainty in the relative positions of the maser spots is estimated
using the expression (Reid et al. \cite{reid})
\begin{equation}
\Delta \theta = \frac{\sigma }{2 \, I } \; {\rm FWHM}
\end{equation}
where FWHM is the un-deconvolved spot diameter, $I$ is the peak intensity and
$\sigma$ is the RMS of the map evaluated over a region where no signal
is present.
Depending on the spot intensity, the relative positional uncertainty varies
in the range $\approx$~0.1--100~$\mu$as.

\section{Results}

\begin{table*}
\begin{center}
\caption{Parameters of the water maser spots in \I.}
\label{tres}
\begin{tabular}{lcccccccc}
\hline 
Number & $\Delta$RA$^{(a)}$ & $\Delta$Dec$^{(a)}$ & $S_\nu$ & $V_{\rm LSR}$ & $V_{\rm RA}$$^{(b)}$ & $V_{\rm Dec}$$^{(b)}$ & $V_{\rm RA}$$^{(c)}$ & $V_{\rm Dec}$$^{(c)}$ \\
 & (mas) & (mas) & (Jy) & (\kms) & (\kms) & (\kms) & (\kms) & (\kms) \\
\hline 
  1 &    0.0$\pm$0.3 &    0.0$\pm$0.3 & 24.30 & --15.0 & --75$\pm$10 &  17$\pm$10 & --84$\pm$10 &  54$\pm$10 \\
  2 &   --1.0$\pm$0.3 &    0.5$\pm$0.3 & 18.36 & --15.3 & --- & --- & --- & --- \\
  3 &   --2.3$\pm$0.3 &    1.2$\pm$0.3 & 10.94 & --16.2 & --- & --- & --- & --- \\
  4 &   --1.6$\pm$0.3 &    0.8$\pm$0.3 &  5.12 & --16.0 & --- & --- & --- & --- \\
  5 &  --24.5$\pm$0.3 &  --51.5$\pm$0.3 &  4.50 &  --2.3 & --39$\pm$11 & --12$\pm$10 & --48$\pm$11 &  25$\pm$10 \\
  6 &  --21.5$\pm$0.3 &  --65.8$\pm$0.3 &  5.58 &  --3.3 & --57$\pm$11 & --23$\pm$11 & --66$\pm$11 &  14$\pm$11 \\
  7 &  --21.1$\pm$0.3 &  --66.6$\pm$0.3 &  0.92 &  --5.1 & --48$\pm$12 & --23$\pm$11 & --58$\pm$12 &  14$\pm$11 \\
  8 &  --22.9$\pm$0.3 &  --63.0$\pm$0.3 &  0.47 &  --5.8 & --42$\pm$11 & --19$\pm$10 & --52$\pm$11 &  19$\pm$10 \\
  9 &  --28.0$\pm$0.3 &  --60.4$\pm$0.3 &  5.77 &  --8.5 & --46$\pm$11 & --20$\pm$11 & --55$\pm$11 &  17$\pm$11 \\
 10 &  --25.8$\pm$0.3 &  --61.6$\pm$0.4 &  2.97 &  --9.5 & --43$\pm$12 & --33$\pm$11 & --53$\pm$12 &   4$\pm$11 \\
 11 &  --26.0$\pm$0.3 &  --60.9$\pm$0.3 &  2.08 &  --8.6 & --43$\pm$11 & --21$\pm$11 & --52$\pm$11 &  16$\pm$11 \\
 12 &   --9.2$\pm$0.3 &    0.1$\pm$0.3 &  0.29 & --14.8 & --67$\pm$11 &  19$\pm$10 & --76$\pm$11 &  56$\pm$10 \\
 13 &   12.3$\pm$0.3 &  --83.8$\pm$0.3 &  0.24 &  --5.6 & --33$\pm$11 & --31$\pm$11 & --42$\pm$11 &   7$\pm$11 \\
 14 &   11.7$\pm$0.3 &  --84.9$\pm$0.3 &  0.08 &  --3.7 & --35$\pm$12 & --20$\pm$11 & --44$\pm$12 &  17$\pm$11 \\
 15 &   14.8$\pm$0.4 &   24.4$\pm$0.4 &  0.49 &  --8.0 & --19$\pm$12 & --19$\pm$11 & --28$\pm$12 &  18$\pm$11 \\
 16 & --141.9$\pm$0.3 &   73.0$\pm$0.3 &  0.85 & --24.9 & --76$\pm$11 &   9$\pm$10 & --85$\pm$11 &  46$\pm$10 \\
 17 & --142.6$\pm$0.3 &   73.2$\pm$0.3 &  0.09 & --24.8 & --- & --- & --- & --- \\
 18 &  --17.8$\pm$0.3 &  --67.0$\pm$0.3 &  0.17 &  --9.5 & --61$\pm$11 & --25$\pm$10 & --70$\pm$11 &  12$\pm$10 \\
 19 &  --19.0$\pm$0.3 &  --66.4$\pm$0.3 &  0.14 &  --7.9 & --49$\pm$11 & --28$\pm$10 & --59$\pm$11 &   9$\pm$10 \\
 20 &    2.1$\pm$0.3 &   --0.7$\pm$0.3 &  0.13 & --14.2 & --66$\pm$11 &  46$\pm$10 & --76$\pm$11 &  83$\pm$10 \\
 21 &    6.4$\pm$0.3 &   29.6$\pm$0.3 &  0.12 &  --8.0 & --- & --- & --- & --- \\
 22 &   29.9$\pm$0.3 &  --77.1$\pm$0.3 &  0.16 &  --3.8 & --- & --- & --- & --- \\
 23 &   28.9$\pm$0.3 &  --78.2$\pm$0.3 &  0.13 &  --3.8 & --- & --- & --- & --- \\
 24 &   21.3$\pm$0.3 &  --87.2$\pm$0.3 &  0.06 &   2.7 & --- & --- & --- & --- \\
 25 &   28.2$\pm$0.3 &  --78.0$\pm$0.3 &  0.10 &  --3.8 & --- & --- & --- & --- \\
 26 &  482.7$\pm$0.3 & --228.3$\pm$0.3 &  1.06 &  --1.9 &   7$\pm$11 & --48$\pm$10 &  --2$\pm$11 & --11$\pm$10 \\
\hline 
\end{tabular}
\end{center}

\vspace*{3mm}
$^{(a)}$~offset with respect to RA(J2000)=20$^{\rm h}$14$^{\rm m}$26\fs0253,
 Dec(J2000)=41\degr13\arcmin32\farcs666 \\
$^{(b)}$~absolute proper motion measured with phase referencing \\
$^{(c)}$~absolute intrinsic proper motion after correction for parallax, motion of the
       Sun with respect to the LSR, and galactic rotation
\end{table*}

\subsection{Maser positions and velocities}

Table~\ref{tres} lists the parameters of the detected maser spots.
Column~1 gives the spot label number. Columns~2 and~3 report the
positional (RA and Dec) offsets (measured on the first epoch of detection)
calculated with respect to the reference spot (labeled \#~1).  Such
offsets are estimated from the (error-weighted) mean
positions of the contributing maser spots. The positional uncertainties are
evaluated by taking the weighted standard deviation of the spot positions.
Columns~4 and~5 list respectively the integrated flux density, $S_{\rm \nu}$,
and the line-of-sight velocity, $V_{\rm LSR}$, of the spot
highest-intensity channel, both averaged over the observational epochs for
the time-persistent spots.

The absolute position of a spot at each observing epoch is calculated
adding the (RA and Dec) offsets to the absolute position of the reference
spot.  The absolute proper motions have been calculated performing a
(error-weighted) linear least-squares fit of the absolute positions with
time.
%
%
Columns~6 and~7 of Table~\ref{tres} report the projected components,
respectively along the RA and Dec axis, of the measured absolute proper
motions, together with the formal errors of the linear least-squares fit.
Using a distance to \I\ of 1.7~kpc, Columns~8 and~9 report the absolute proper
motion components after correction for parallax, motion of the Sun with
respect to the LSR, and galactic rotation.

\subsection{The distance to \I}
\label{sdist}

The absolute proper motions measured by us are the sum of the intrinsic
proper motion due to the velocity of the maser spots with respect to the YSO
and the apparent motion due to the annual parallax and motion of the
source with respect to the Sun. Therefore, one of the possible outcomes of a
proper motion study is the derivation of the distance to the source by
fitting the annual parallax. However this technique cannot be applied in our
case, as the few measurements (3) and the short intervals between them
($\le$95~days) are insufficient to properly sample the apparent position of
the spots as a function of time -- at least 5 points evenly distributed over 1
year would be needed (see e.g. Vlemmings et al. \cite{vlem}). Also, for the
\WAT\ masers in \I\ the situation is complicated by the fact that the
intrinsic proper motion of each spot (4--14~mas\,yr$^{-1}$)
is greater (see Sect.~\ref{sres}) than that due to the annual parallax
($\la$1.2~mas\,yr$^{-1}$).

The same considerations hold when deriving the intrinsic proper motion of
each spot: the apparent proper motion due to the parallax and
source velocity must be subtracted\footnote{
The contribution of the parallax and source motion relative to the Sun have
been calculated with a program kindly provided by Prof.~Tetsuo~Sasao.
Both a flat galactic rotation curve and the rotation curve of Brand \& Blitz
(\cite{brbl}) have been considered, but the results obtained for a distance
of 1.7~kpc were not significantly different.} from the observed proper motion.
Such an apparent motion depends on the distance and consequently also the
estimate of the intrinsic motion depends on it. One may thus use this
dependency to constrain the distance. Two approaches are possible.

\begin{figure}
\includegraphics[angle=-90,width=8.5cm]{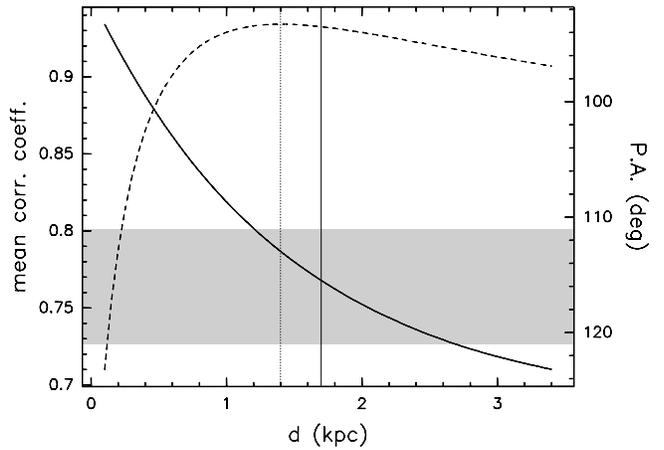}
\caption{
 Plot of mean correlation coefficient of the maser spots (dashed curve;
 see text) and mean direction of the normalised proper motions (solid curve)
 versus source distance. The vertical dotted and solid lines mark respectively
 $d$=1.4 and 1.7~kpc. The grey area denotes the range of possible jet position
 angles.
}
\label{fccphi}
\end{figure}

The first uses the fact that, to a first order approximation, the maser spots
should move along a given direction, since they are participating in the
expansion of the jet from \I. This implies that for each spot the positions
on the plane of the sky at the three epochs should lie along a straight
line. A quantitative estimate of this is obtained from the correlation
coefficient, $r$, of each triplet of positions per spot. We have hence
computed $r$ for all spots and calculated the arithmetic mean. This is shown
in Fig.~\ref{fccphi} as a function of the distance, $d$. A maximum is reached
for $d$=1.4~kpc, but also larger distances cannot be excluded, whereas for
$d<1$~kpc the correlation coefficient drops significantly.

An alternative approach is that of comparing the mean direction of the
proper motions (after subtraction of the apparent motion) with the direction
of the jet axis. To give equal weights to all spots we have performed a 
vector average of the unity vectors of the proper motions in columns~6 and~7
of Table~\ref{tres}, weighted by the corresponding errors. The result is
shown in Fig.~\ref{fccphi} as a function of the distance. The grey area in
the same figure marks the range of plausible values of the jet position angle
(P.A.), obtained from tracers such as the SiO(2--1) and \HM\ lines (Cesaroni et
al.  \cite{cesa99}), and the 2.2~\mic\ (Edris et al. \cite{edris},
hereafter EFCE) and 3.6~cm continuum emission (Hofner et al. \cite{hofner};
Hofner pers. comm.). Clearly, any distance between 1.2 and 2.7~kpc is
possible.

We conclude that with the present data only loose constraints can be set on the
distance to \I, indicating that the source cannot be closer than
1.2~kpc. In the following we will assume the usual estimate of 1.7~kpc
for consistency with previous studies of the source.

\subsection{Distribution and velocities of the \WAT\ maser spots}
\label{sres}

\begin{figure*}
\includegraphics[angle=0,width=15.5cm]{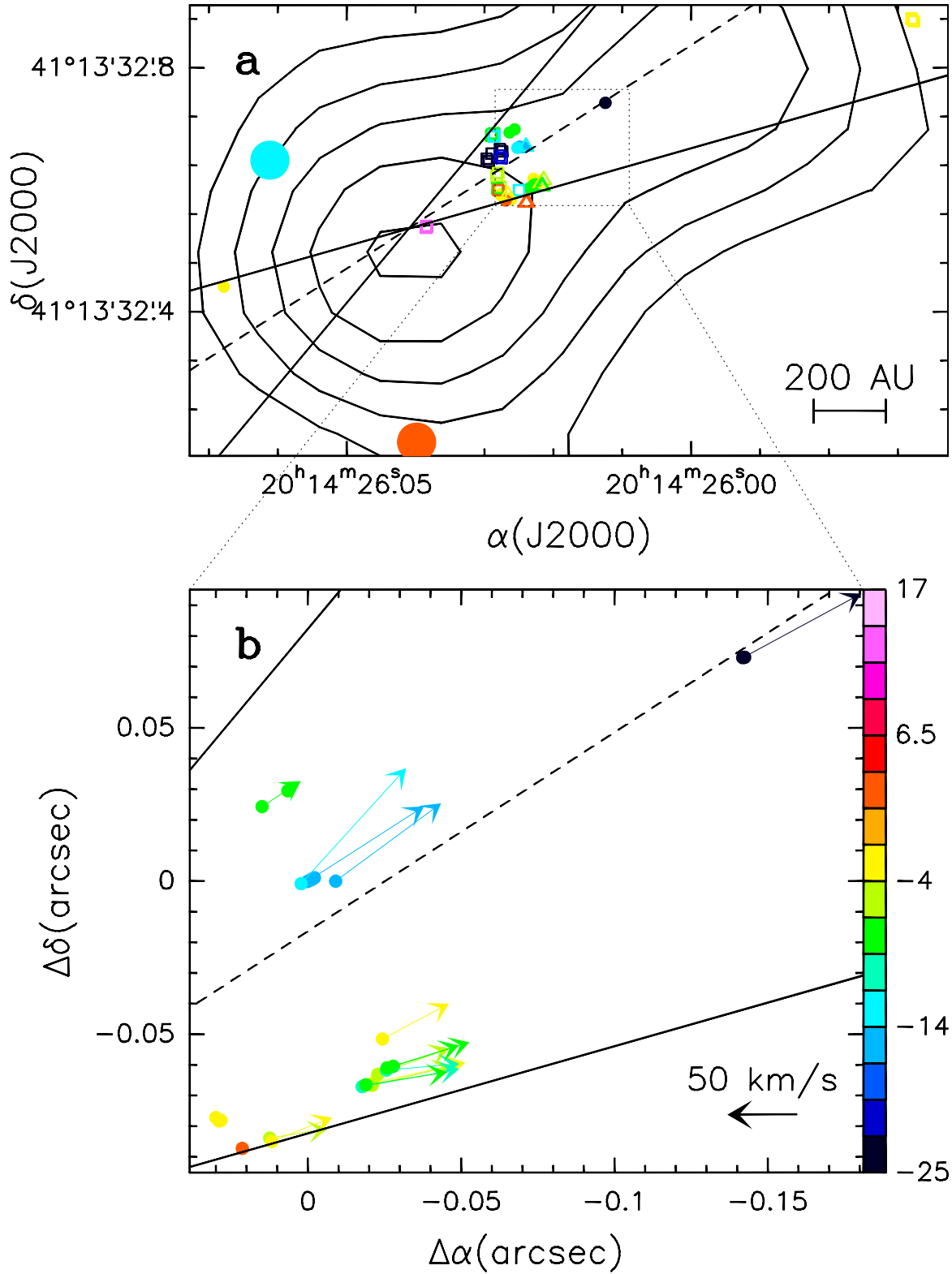}
\caption{
 {\bf a)}
 Map of the \WAT\ maser spots detected towards \I\ by us (small circles), MCR
 (squares), and EFCE (triangles) overlayed on a
 contour map of the 3.6~cm continuum emission (Hofner et al.  \cite{hofner}).
 Also shown are the OH maser emission peaks observed by EFCE (big circles).
 The errors on the {\it absolute} positions of
 the \WAT\ maser spots detected by us are given in Table~\ref{tres}, while
 those of MCR and EFCE are respectively 30~mas and
 15~mas; for the OH masers the error is 25~mas. The colour denotes the LSR
 velocity of each spot according to the colour scale in the bottom panel.
 The solid lines indicate the conical jet which represents the best fit to
 the \WAT\ maser spots obtained with the model discussed in Sect.~\ref{smod}.
 {\bf b)}
 Enlargement of the central region illustrating the locations and absolute
 proper motions (corrected for parallax, solar motion with respect to the LSR,
 and galactic rotation) of the
 \WAT\ maser spots detected in this study. Offsets in RA and Dec are measured
 with respect to RA(J2000)=20$^{\rm h}$14$^{\rm m}$26\fs0253,
 Dec(J2000)=41\degr13\arcmin32\farcs666. No proper motion measurement is
 available for points without an associated arrow. The colour coding is the
 same as for the top panel. Note that the systemic LSR velocity of \I\ is
 --3.5~\kms.
}
\label{fpmots}
\end{figure*}

As previously mentioned, the main scope of our study was to investigate the
geometry and kinematics of the jet in \I\ using the \WAT\ maser spots as test
particles of the outflowing molecular gas. Figure~\ref{fpmots}a shows all
the \WAT\ maser spots detected in the region by us, MCR, and EFCE. Also shown
are the OH maser emission peaks, which according to EFCE mark the plane of
the circumstellar disk. We overlay the masers on a contour map of the 3.6~cm
continuum emission imaged by Hofner et al. (\cite{hofner}), corresponding to
the inner ionised part of the jet/outflow in \I. All of these tracers agree
very well with the pattern of the conical jet model that will be discussed in
Sect.~\ref{smod} and is already presented here for the sake of comparison. As
expected, the vertex of the cone -- i.e. the putative position of the YSO
powering the jet -- falls very close to the peak of the 3.6~cm continuum
emission and in between the two OH maser peaks denoting the disk.

All this strongly favours the scenario already outlined by MCR, according
to which the \WAT\ masers are located at the interface between a conical jet
and the surrounding quiescent material, and are flowing along the surface of
such a jet with velocities directed outward from the central YSO. Such an
interpretation is fully confirmed by the proper motions of the \WAT\ maser
spots, shown in Fig.~\ref{fpmots}b: clearly, the spot velocities are
diverging from the YSO. We stress again that in order to obtain the intrinsic
proper motions of the spots (columns~8 and~9 of Table~\ref{tres}) the
absolute proper motions (columns~6 and~7) have been
corrected for the annual parallax and motion of the source relative to the
Sun, assuming a distance of 1.7~kpc.

\begin{figure}
\includegraphics[angle=0,width=8.5cm]{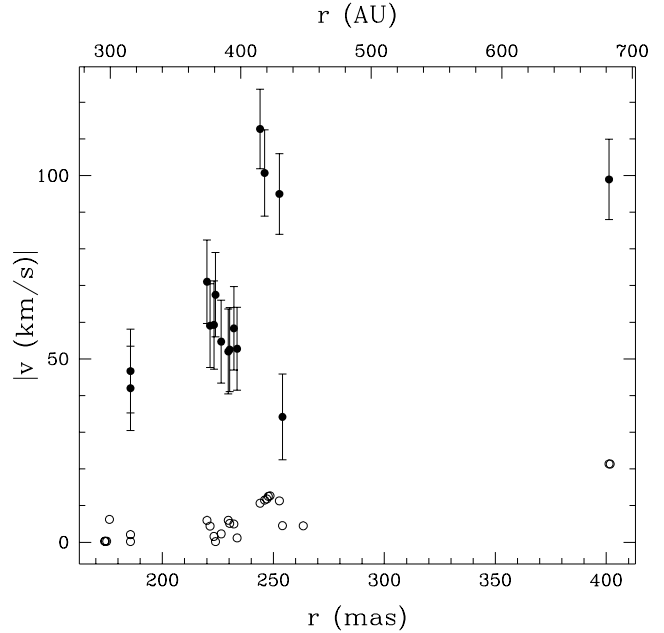}
\caption{
 Plot of speed versus separation (measured on the plane of the sky) from the
 3.6~cm continuum peak for all \WAT\ maser spots of Fig.~\ref{fpmots}b with
 measured proper motions (filled circles). Also shown is a plot of the l.o.s.
 component of the velocity for all spots detected in this study (empty
 circles). Note the tendency of the speed to increase with distance from the
 YSO.
}
\label{fvr}
\end{figure}

Combining the proper motions with the LSR velocity of each spot after
subtraction of
the systemic velocity of \mbox{--3.5~\kms},
one can obtain the corresponding speed of each spot. In Fig.~\ref{fvr} we
plot this (filled circles) as a function of the distance $r$ (measured on the
plane of the sky) from the peak of the 3.6~cm continuum emission, which is a
good guess for the location of the YSO. Although with a large scatter,
the speed tends to increase with $r$. A similar trend is seen for the l.o.s.
components of the spot velocities -- which represent a lower limit to the full
speed. These are shown in the same figure (empty circles) also for the spots
that do not have measured proper motions. We conclude that the jet is
accelerated within at least 700~au from the YSO.

For a more quantitative description of the maser distribution and velocities,
in the next section we elaborate a simple model which represents an improvement
on the model developed by MCR.

\section{A model for the \WAT\ masers in \I}
\label{smod}

The scope of the jet model presented here is to prove that the scenario
outlined in the previous section for the origin of the \WAT\ maser emission
in \I\ is correct. Unlike MCR's model, ours
takes into account that the spot speed increases with distance from the YSO
and fits also the velocity components in the plane of the sky.
We assume that the maser spots lie on the surface of a cone with
semi-opening angle $\theta$, inclination angle with respect to the l.o.s.
$\psi$, and vertex coincident with the YSO at \az,\dz. The spot velocities are
directed radially outward from the vertex and are proportional to the
distance, $R$, from it: $v(R)=\dvdr\,R$. The coordinate system is centred on
the vertex, with $z$ along the l.o.s. (the observer lies at $z=-\infty$) and
$x$ coincident with the projection of the jet axis on the plane of the sky.
For further details and a sketch of the model we refer to Sect.~4.1 and
Fig.~4 of MCR.

Under the previous assumptions one can express the components of the velocity
along the axes as
\begin{eqnarray}
v_{\rm x} & = & \dvdr \, x        \label{evx} \\
v_{\rm y} & = & \dvdr \, y        \label{evy} \\
v_{\rm z} & = & \dvdr \, z        \label{evz}
\end{eqnarray}
where \dvdr\ is constant and
\begin{eqnarray}
z & = & \left[ \rule{0cm}{5mm} x \S \C \, (1+\Gq) \right. \nonumber \\
  & & \left. \pm \G \sqrt{x^2+y^2(\Cq-\Gq\Sq)} \right] \times  \label{ezeta} \\
  & & \times \left( \Gq\Sq - \Cq \right)^{-1}. \nonumber
\end{eqnarray}
Note that Eq.~(2) of MCR contains an unfortunate typing error ($\tan^2\theta$
must be replaced by $\cot^2\theta$) which makes it different from our
Eq.~(\ref{ezeta}). The correct equation was used in the calculations of
MCR.

The input parameters of the model are: the P.A. of the jet, $\theta$, $\psi$,
\dvdr, and the coordinates of the cone vertex, \az\ and \dz. We searched for
the values of these parameters minimising the expression
\begin{equation}
 \chi^2 = \sum_{i=1}^{26}
         \left(v_{\rm x}^{(i)} - V_{\rm x}^{(i)}\right)^2 +
         \left(v_{\rm y}^{(i)} - V_{\rm y}^{(i)}\right)^2 +
         \left(v_{\rm z}^{(i)} - V_{\rm z}^{(i)}\right)^2
\end{equation}
where $i$ indicates the spot number in Table~\ref{tres} and symbols $v$ and
$V$ denote respectively the model and observed velocities; for spots without
measured proper motions only the term involving the $z$ components is considered
in the summation.

\begin{figure}
\includegraphics[angle=0,width=8.5cm]{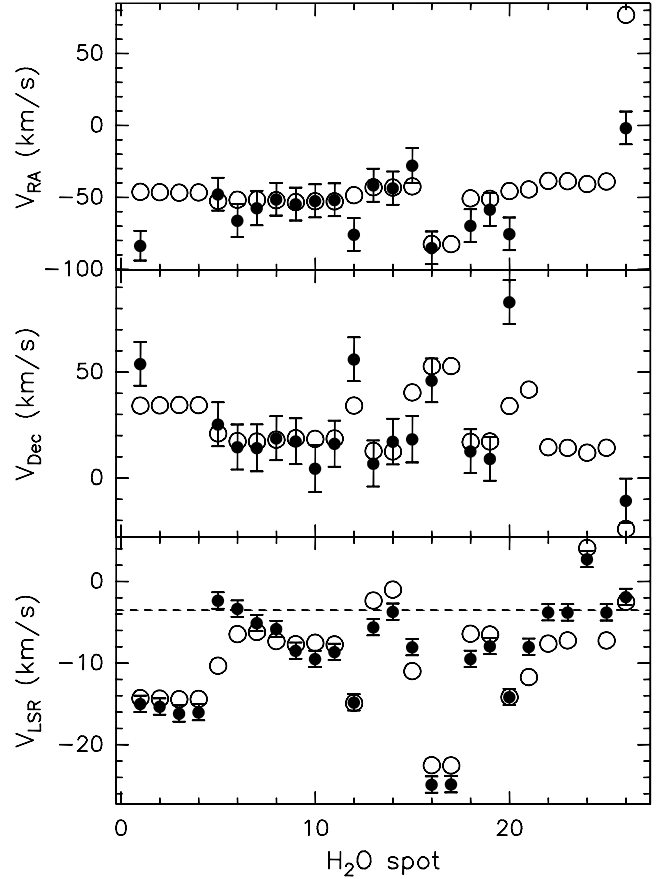}
\caption{
 Comparison between the observed velocities (filled circles) and those obtained
 from our model fit (empty circles). The top and middle panel show respectively
 the proper motions in the RA and Dec directions, while the bottom panel
 plots the LSR velocity. Note that data points missing in the top and middle
 panels correspond to spots without measured proper motions. The dashed line
 marks the systemic velocity (--3.5~\kms) of \I. The number on the x-axis
 indicates the maser spot according to the notation in Table~\ref{sres}.
}
\label{ffit}
\end{figure}

The best fit is obtained for P.A.=123\degr, $\psi$=96\degr, $\theta$=17\degr,
\az(J2000)=20$^{\rm h}$14$^{\rm m}$26\fs0410,
\dz(J2000)=41\degr13\arcmin32\farcs536, and
\dvdr=0.255~\kms\,mas$^{-1}$=0.150~\kms\,au$^{-1}$.
Figure~\ref{ffit} compares the measured values of the proper motions and LSR
velocities (filled circles) with those obtained from the model fit (empty
circles). Clearly the agreement is very good, proving that our model
provides a satisfactory interpretation for the origin of the water masers in \I.

A comparison between our best fit parameters and those of MCR reveals
significant differences, the most striking being the different expansion
velocities: 23~\kms\ in MCR and 34--112~\kms\ in our case (see
Fig.~\ref{fvr}). This is due to the fact that the spots are moving almost in
the plane of the sky, so that it is very difficult to correct for the
inclination angle without proper motion information. As a matter of fact,
$\psi$ is larger in our model than in MCR. Moreover, our value
($\psi$=96\degr) suggests that the north-western lobe of the jet is pointing
away from us, whereas in MCR ($\psi$=59\degr) the opposite occurs. Our result
is consistent with recent findings: on the one hand, the jet is very close to
the plane of the sky on scales as large as 10\arcsec, or 0.08~pc (Cesaroni et
al. \cite{cesa99}); on the other hand, the jet axis is known to undergo
precession, with the NW lobe receding from the observer at the present time
(Cesaroni et al. \cite{cesa05}). It is hence not surprising that on scales as
small as a few 100~au the jet has already crossed the plane of the sky, as
indicated by the value $\psi$=96\degr.

As for the opening angle of the cone, MCR's estimate ($\theta$=29\degr) is
twice as much as ours (17\degr). This is partly related to the fact that
also the vertex of the cone has moved by $\sim$84~mas to the SE with respect
to MCR's estimate. As discussed in Sect.~\ref{sres} the new position is
closer to the disk plane, identified by the OH maser emission
(see Fig.~\ref{fpmots}a).

In conclusion, we believe that the estimates of the \WAT\ maser jet
obtained in the present study represent an improvement on those of MCR
and are hence to be preferred.

\section{Variability of the \WAT\ masers}

\begin{figure*}
\includegraphics[angle=-90,width=18cm]{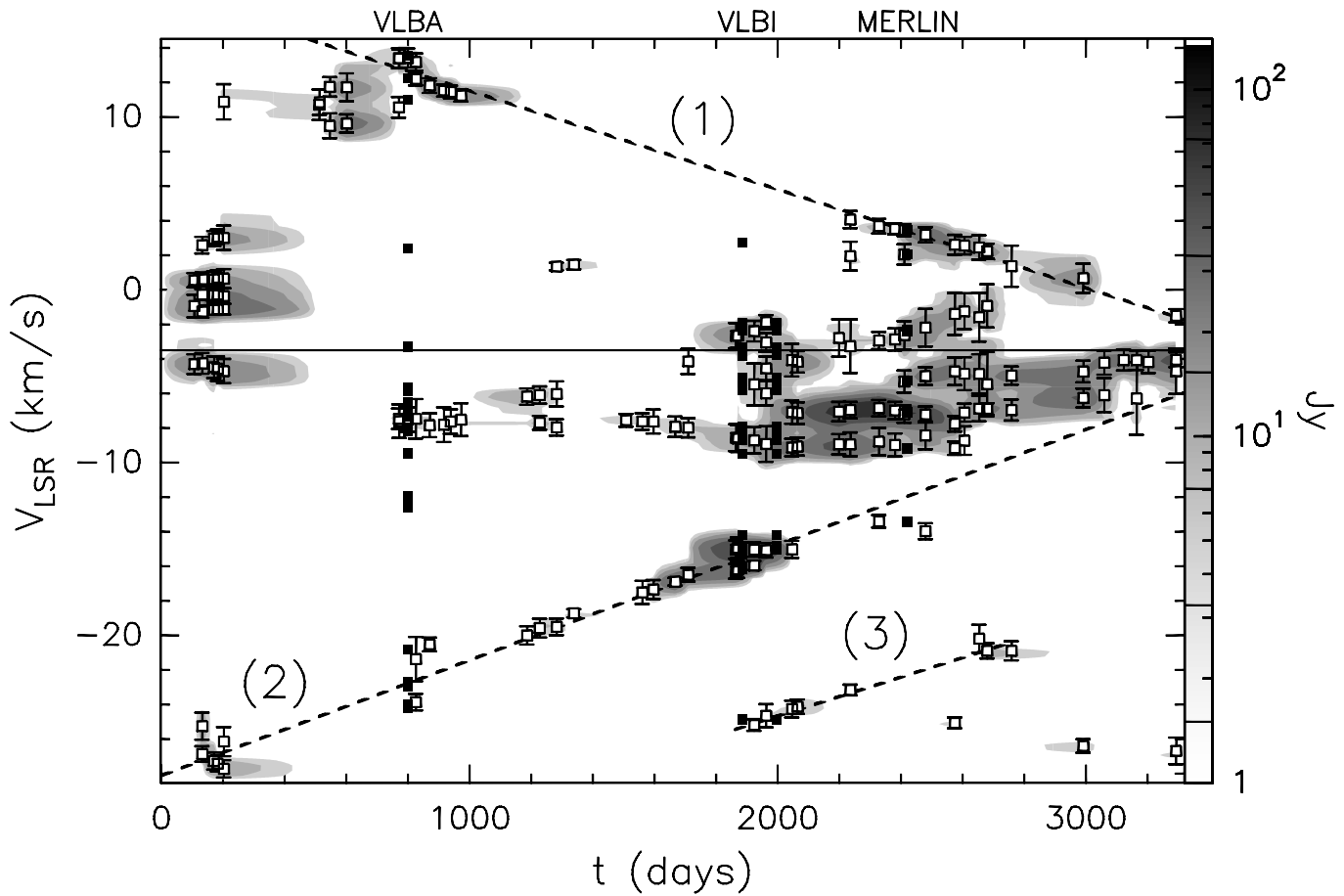}
\caption{
 Variability plot of the \WAT\ maser emission in \I. The grey scale map
 represents the intensity measured with the Medicina antenna as a function
 of time and LSR velocity. The value $t$=0 corresponds to September 12, 1995.
 The white squares mark the positions of the the peak velocities of the lines
 detected in the Medicina spectra, while the error bars correspond to the
 line full width at half maximum (FWHM). Both the peak velocity and line width
 have been obtained with a Gaussian fit. The black squares indicate the
 peak velocities of the spots observed in the VLBA, VLBI, and MERLIN
 observations of MCR, this study, and EFCE respectively. Only the first and
 third VLBI epochs are shown here to preserve readability of the figure. The
 horizontal solid line marks the systemic velocity (--3.5~\kms) of \I. The
 dashed lines outline the linear variation of velocity versus time observed
 in three spectral features, labeled with numbers (1) to (3).
}
\label{fvar}
\end{figure*}

It is interesting to complement the high resolution study of the spatial
distribution of the \WAT\ masers in \I\ with an analysis of their time
variability. Water masers are known to be highly variable and \I\ is no
exception to this rule. Since 1987, the Medicina 32-m
antenna\footnote{
The Medicina VLBI radiotelescope is operated by the INAF-Istituto di Radioastronomia.
}
has been used as a single-dish telescope
to monitor a number of galactic \WAT\ masers with typical
sampling intervals of 2--3 months. For details on the observations and data
analysis we refer to Valdettaro et al. (\cite{valde}). \I\ belongs to this
sample, so that we could retrieve the \WAT\ spectra from the Medicina
database and analyse them with the method developed by Valdettaro et al.
(\cite{valde}). This is shown in Fig.~\ref{fvar}, which represents an image
of the maser intensity as a function of time and LSR velocity. Also shown are
the points corresponding to the peak velocities of the \WAT\ lines
detected at different epochs, including those obtained by us with global
VLBI, by MCR with the VLBA, and by EFCE with MERLIN.

The most striking characteristic is the presence of three clear velocity
trends, outlined by the dashed lines in Fig.~\ref{fvar}: in all cases, the
velocity decreases linearly with time, with slopes $\d V_{\rm LSR}/\d t\simeq
-2.1, -2.7,$ and --2.1~\kms\,yr$^{-1}$ for features (1), (2), and (3)
respectively. In the following we discuss these features in detail.

\subsection{Features (2) and (3)}

For features (2) and (3) one may easily derive the value of the deceleration
corrected for the inclination angle. In fact, we have measured the proper
motion of the corresponding spots: using the notation of Table~\ref{tres},
these are spots 1, 12, and 20 for feature (2) and 16 for feature (3). The
inclination angle is given by the ratio between the l.o.s. velocity and the
projection of the velocity on
the plane of the sky. Applying this correction, one obtains
$\d V/\d t\simeq-18$~\kms\,yr$^{-1}$ for (2) and --9.7~\kms\,yr$^{-1}$ for (3).

\begin{figure}
\includegraphics[angle=-90,width=8.5cm]{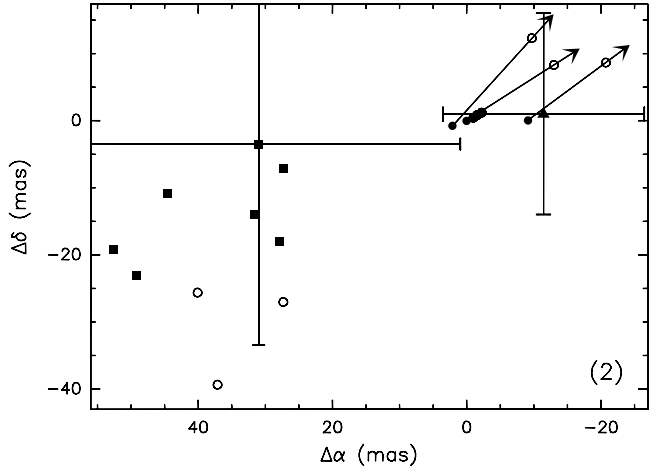}
\caption{
 Map of the \WAT\ maser spots belonging to feature (2) in Fig.~\ref{fvar}.
 The (0,0) position corresponds to RA(J2000)=20$^{\rm h}$14$^{\rm m}$26\fs0253,
 Dec(J2000)=41\degr13\arcmin32\farcs666.
 Black circles indicate our measurements, squares those by MCR, and triangles
 those by EFCE. The error bars indicate the 1$\sigma$ uncertainty
 in the astrometry of the MCR and EFCE measurements (the error on our data
 is less than the size of the symbols). Note that the error in the
 {\it relative} positions of the spots in each data set is of order 1~mas.
 The arrows denote the proper motions measured by us, while the empty
 circles mark the positions of the same spots extrapolated to the
 dates of MCR (November 21, 1997) and EFCE's (March 2002)
 observations, on the basis of the current spot position, proper motion, and
 acceleration (see text).
}
\label{fgr2}
\end{figure}

In Fig.~\ref{fgr2} the distribution of the spots corresponding to feature (2)
is shown, including our measurements and those of MCR and EFCE.
Notwithstanding the large positional errors on the VLBA and MERLIN
measurements, the general trend follows the direction of the \WAT\ maser jet
depicted in our model. This is proved by the empty circles, which mark the
positions of the spots with measured proper motions (filled circles with
arrows) extrapolated at the two epochs of the VLBA and MERLIN observations.
Such an extrapolation takes into account the acceleration in the plane of the
sky estimated above. Clearly, the location of the empty circles is consistent
within the errors with that of the VLBA (squares) and MERLIN (triangles)
spots.

It is worth noting that the error bars in Fig.~\ref{fgr2} represent the
uncertainty on the {\it absolute} position at each epoch. The {\it relative}
position of the spots observed at a given epoch is instead very accurate,
of order $\sim$1~mas. Therefore, it is clear that in the first epoch
(squares) the dispersion between the spots was much greater than at the
time of our VLBI observations. One may speculate that this an indication
of the fact that the shocks associated with the masers are converging.


\subsection{Feature (1)}

\begin{figure}
\includegraphics[angle=-90,width=8.5cm]{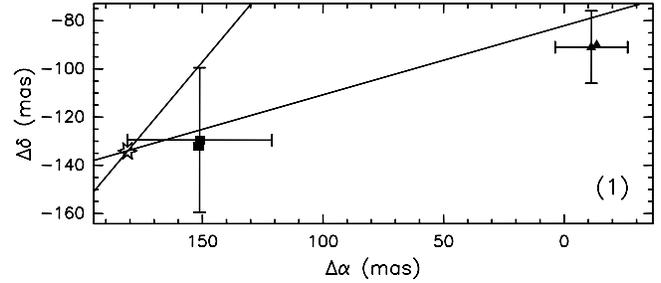}
\caption{
 Map of the \WAT\ maser spots belonging to feature (1) in Fig.~\ref{fvar}.
 The (0,0) position corresponds to RA(J2000)=20$^{\rm h}$14$^{\rm m}$26\fs0253,
 Dec(J2000)=41\degr13\arcmin32\farcs666. Squares indicate measurements by
 MCR and triangles those by EFCE. The error bars indicate the 1$\sigma$
 uncertainty on the astrometry. The error on the {\it relative} position of
 the spots in each data set is of order 1~mas. The starred polygon marks the
 position of the YSO according to our conical model fit, while the solid
 lines denote the projection of the cone on the plane of the sky.
}
\label{fgr1}
\end{figure}

Feature (1) has not been detected in our VLBI observations, so that no proper
motion measurement is available. However, the VLBA and MERLIN measurements
of MCR and EFCE can be used to obtain an estimate of the average proper
motion. Notwithstanding the large uncertainty on the absolute
positions of the spots, the time interval between the two measurements is
large enough to guarantee an angular separation between the positions at the
two epochs much larger than the errors. This can be seen in Fig.~\ref{fgr1},
where also a sketch of the conical model presented in Sect.~\ref{smod} is
shown for the sake of comparison. The mean velocity in the plane of the sky
is obtained from the separation between the two positions (169$\pm$45~mas or
287$\pm$77~au) divided by the corresponding time interval ($\sim$1620~days),
and turns out to be 306$\pm$42~\kms. We also stress that the direction of the
proper motion is perfectly consistent with expansion along the cone as
expected in our model: this has the twofold consequence of confirming our
model and proving that all spectral features falling on the same dashed line
in Fig.~\ref{fvar} do arise from a given spot (or group of spots) moving
along the jet.

One may compute the mean velocity of feature (1) along the l.o.s. from
the arithmetic mean between the l.o.s. velocities at the two epochs
($\sim$11.2~\kms). Finally, from the ratio between this and the proper
motion, it is possible to obtain the inclination of the velocity vector with
respect to the l.o.s. and thus correct the l.o.s. values of the mean speed
(11.4~\kms) and acceleration (--2.1~\kms\,yr$^{-1}$): the corrected values
are 306~\kms\ and --56~\kms~yr$^{-1}$.

\subsection{Origin of the deceleration}

From the previous findings, we conclude that \WAT\ masers in \I\ are
undergoing a strong deceleration ranging from --50 to --10~\kms~yr$^{-1}$.
Although evidence of deceleration from single-dish monitoring of \WAT\ masers
has already been reported by other authors (see Brand et al. \cite{brand}),
to our knowledge this is the first time that such a finding is confirmed by
high-angular resolution observations.

This existence of deceleration may look contradictory with the positive
velocity trend shown in Fig~\ref{fvr}. However, $\d V/\d R>0$ does not
necessarily imply $\d V/\d t>0$, because the $V(R)$ provided by our VLBI
measurements corresponds to an {\it instantaneous picture} of the spots
velocity field, whereas $V(t)$ obtained from the variability analysis
describes the {\it evolution of each single spot} with time.

In order to reconcile $\d V/\d R>0$ with $\d V/\d t<0$, we propose the
following scenario. Water masers are very likely generated in the post-shock
region of J-type shocks (Elitzur et al. \cite{elhomc}). In our model,
such shocks occur along the conical surface representing the interaction
between the jet and the surrounding cloud. As predicted by theoretical
models (e.g. Ferreira \& Casse \cite{ferr}, Shu et al. \cite{shu}),
the jet should be accelerated close to the YSO. Consequently, the further
away (from the YSO)
the jet impinges against the surrounding material, the larger will be
the velocity of the corresponding shock. At this time a maser spot ``is
switched on'', with the same velocity as the shock.  This explains why
$\d V/\d R>0$. After that, the shock is slowed down as it proceeds through
the dense material of the cloud, thus causing the observed deceleration,
$\d V/\d t<0$, of the spot velocity.

\begin{figure}
\includegraphics[angle=-90,width=8.5cm]{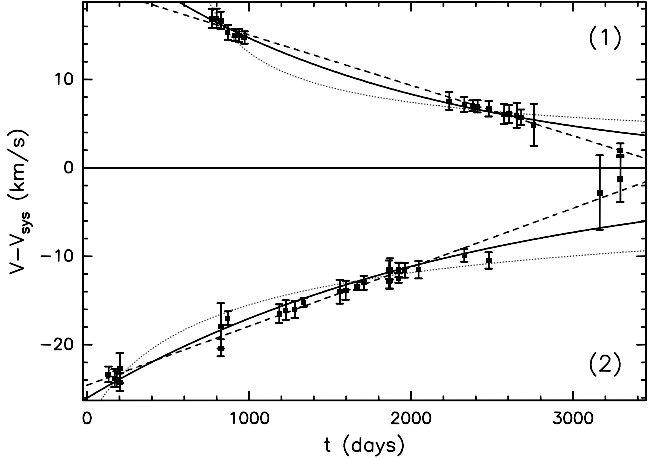}
\caption{
 Peak velocity of feature (1) and (2) (see Fig.~\ref{fvar}) as a function of
 time. Error bars denote line FWHMs. The dashed lines are the same as in
 Fig.~\ref{fvar} and represent linear fits to the data, while dotted and
 solid lines are fits of the type $V\propto (t-t_0)^{-1/2}$ and
 $V\propto \exp[-(t-t_0)/\tau]$, respectively.
}
\label{fvt}
\end{figure}

It remains to be verified that the scenario depicted above can fit the
observed trend of $V(t)$ quantitatively.
If the deceleration is due to braking of a shock,
this should imply a power-law for $V(t)$. In fact, for shocks conserving
energy one has $MV^2$=const. $\Rightarrow$ $Vt\,V^2$=const. $\Rightarrow$
$V\propto t^{-1/3}$, while shocks conserving momentum imply $MV$=const. and
hence $V\propto t^{-1/2}$. Another possibility is that of an exponential
decay of the shock velocity: na\"{\i}vely, this may be described with a sort
of ``viscosity'' in the medium ahead of the shock.
With this in mind, we have fitted the velocity trends of features (1) and (2)
with $V\propto (t-t_0)^{-1/2}$ and $V\propto \exp[-(t-t_0)/\tau]$: these are
represented respectively by the dotted and solid curves in Fig.~\ref{fvt},
where we plot the l.o.s. velocities of features (1) and (2) as a function of
time. Also shown are linear fits to the data (dashed lines). It is worth
noting that in all cases the lifetime of the maser spots ranges between 5 and
12~yr, a timescale comparable to our monitoring period. Clearly, it is
difficult to rule out any of the models, although the exponential fit seems
more satisfactory. A more sensitive and frequent monitoring of the
\WAT\ maser emission in \I\ is necessary to discriminate between models.

\section{Summary and conclusions}

We have performed three-epoch VLBI observations of the \WAT\ maser emission
from the massive protostar \I. Phase referencing allows us to measure the
{\it absolute} proper motions of the maser spots and hence obtain a 3-D
picture of the velocity field. These high angular resolution results have
been complemented by an analysis of the variability of the \WAT\ maser emission
through single-dish observations spread over 15 years. The main results obtained
from this study and comparison with previous data by MCR and EFCE are the
following:
\begin{itemize}
\item
 A total of 26 spots have been detected in our VLBI observations, 17 of which
 at all three epochs thus allowing measurement of their proper motions.
\item
 An attempt
 to use proper motions to obtain an estimate of the distance to the source
 may only loosely constrain it between 1.2 and 2.7~kpc.
\item
 After correcting the measured proper motions for annual parallax and
 apparent motion of the source with respect to the Sun (assuming a distance of
 1.7~kpc), we find that the maser spots are expanding from
 a common origin, consistent with the expected position of the protostar.
 The speed is increasing from 34 to 112~\kms\ for increasing distance from
 the protostar.
\item
 The l.o.s. velocities and proper motions of the spots can be fitted with an
 improved version of the conical jet model proposed by MCR in a previous VLBA
 study of the same maser source. The best fit is obtained for a jet
 semi-opening angle of 17\degr, an inclination of 96\degr\ with respect to
 the l.o.s. (the jet axis to the NE is pointing away from the observer),
 an expansion velocity gradient of 255~\kms\,yr$^{-1}$, and a cone vertex
 located at RA(J2000)=20$^{\rm h}$14$^{\rm m}$26\fs0410,
 Dec(J2000)=41\degr13\arcmin32\farcs536. This is consistent with the
 expected position of the protostar.
\item
 The variability study demonstrates that some of the maser features undergo
 decelerations from $\sim$~--50 to --10~\kms\,yr$^{-1}$, with lifetimes of
 the order 5--12~yr. We speculate that this may be due to braking of the shocks
 where the maser emission is originating, at the interface between an
 accelerated jet and the surrounding molecular environment.
\end{itemize}

We conclude that our study has shed light on the 3-D structure and kinematics
of the jet in \I\ and outlined some intriguing characteristics of the
time variability of the maser -- and hence of the jet. A full understanding
of this latter aspect will require more sensitive observations with better
angular and, possibly, temporal resolution.

\begin{acknowledgements}
It is a pleasure to thank Prof. Tetsuo Sasao for allowing us to use his
program for the computation of apparent proper motions due to annual
parallax, solar motion with respect to the LSR, and galactic rotation. We are
also grateful to Gary Fuller for making available the results of his
MERLIN observations,
prior to publication. Many thanks are due to Francesco Palla and Jan Brand
for a
critical reading of the manuscript and Daniele Galli and Rino Bandiera for
stimulating discussions.
\end{acknowledgements}

\end{document}